\documentclass[superscriptaddress,showpacs,floatfix]{cpbtex}

\usepackage[ruled, linesnumbered]{algorithm2e}
\usepackage[title]{appendix}
\usepackage{booktabs}
\usepackage{braket}
\usepackage{cleveref}
\usepackage{xcolor}
\usepackage{listings}
\usepackage{graphicx}
\usepackage{makecell}

\crefname{figure}{Fig.}{Figs.}
\crefname{table}{Table}{Tables}

\begin{document}

\title{Practical algorithm for simulating thermal pure quantum states}

\author{Wei-Bo He$^{1, 2}$, \ Yun-Tong Yang$^{1, 2}$ \ and \ Hong-Gang Luo$^{1, 2,}$\thanks{Corresponding author. E-mail:luohg@lzu.edu.cn}\\
    $^{1}${\small School of Physical Science and Technology, Lanzhou University, Lanzhou 730000, China} \\  
    $^{2}${\small Lanzhou Center for Theoretical Physics, Key Laboratory of Theoretical Physics of Gansu Province,} \\
          {\small Key Laboratory of Quantum Theory and Applications of MoE,} \\
          {\small Gansu Provincial Research Center for Basic Disciplines of Quantum Physics,} \\
          {\small Lanzhou University, Lanzhou 730000, China} \\
}

\date{\today}
\maketitle

\begin{abstract}
    The development of novel quantum many-body computational algorithms relies on robust benchmarking. However, generating such benchmarks is often hindered by the massive computational resources required for exact diagonalization or quantum Monte Carlo simulations, particularly at finite temperatures. In this work, we propose a new algorithm for obtaining thermal pure quantum states, which allows efficient computation of both mechanical and thermodynamic properties at finite temperatures. We implement this algorithm in our open-source C++ template library, \textit{Physica}. Combining the improved algorithm with state-of-the-art software engineering, our implementation achieves high performance and numerical stability. As an example, we demonstrate that for the $4 \times 4$ Hubbard model, our method runs approximately $10^3$ times faster than $\mathcal{H}\Phi$ 3.5.2. Moreover, the accessible temperature range is extended down to $\beta = 32$ across arbitrary doping levels. These advances significantly push forward the frontiers of benchmarking for quantum many-body systems.
\end{abstract}

\textbf{Keywords:} \textit{Physica}, thermal pure quantum states, hubbard model, strong correlated electron systems

\textbf{PACS:} 01.50.hv, 02.70.-c, 05.30.-d, 71.10.-w

\section{Introduction}
    Quantum many-body physics stands as one of the most challenging and fascinating areas of modern physics. In particular, the complex and rich emergent phases in strongly correlated systems often elude description by conventional frameworks such as Landau Fermi liquid theory and the Landau-Ginzburg paradigm. Alongside physical experiments and quantum simulations, numerical methods offer a powerful avenue for validating theoretical predictions. Although a variety of computational techniques exist—including exact diagonalization (ED), quantum Monte Carlo (QMC) \cite{DQMC, SSE, CTQMC}, density matrix renormalization group (DMRG) \cite{DMRG1, DMRG2}, tensor networks (TN) \cite{TN}, and dynamical mean field theory (DMFT) \cite{DMFT1, DMFT2, DMFT3}—only sign-problem-free QMC and ED are generally considered reliable enough for benchmarking. However, these two face significant limitations: for instance, at arbitrary temperature the Hubbard model is sign-problem-free only at half-filling, which excludes much of the experimentally relevant doping regime—including potential superconducting phases. While ED is primarily a ground-state method, extensions such as the finite-temperature Lanczos method (FTLM) \cite{FTLM1, FTLM2} have been developed to approximate finite-temperature properties in small systems. Although FTLM can produce accurate results with a reduced number of Lanczos steps and randomized state sampling, it requires careful balancing between computational cost and numerical accuracy, including the trade-off between sufficient step count and expensive reorthogonalization procedures \cite{FTLM1}.

    The open-source community plays a vital role in advancing research by providing fast and reliable code for reproducible benchmarking. In recent years, several open-source software packages for quantum many-body computations have been developed. For example, ALF \cite{ALF1, ALF2} and hubbard-DQMC \cite{hubbard-DQMC} are QMC packages targeting fermionic systems; however, both are significantly affected by the notorious sign problem, especially at low temperatures. On the other hand, QuSpin \cite{QuSpin} offers a user-friendly Python environment for ED and dynamics simulations, but limiting to small and medium-sized quantum systems. XDiag \cite{XDiag}, a recently developed ED library written in C++ with a Julia interface, strikes a balance between computational performance and usability. Meanwhile, $\mathcal{H}\Phi$ \cite{HPhi} delivers an ED package optimized for large-scale distributed-memory clusters and supports finite-temperature calculations via thermal pure quantum (TPQ) states.

    In the present work, we focus on the TPQ technique. The TPQ states provide a powerful approach for approximating various mechanical and thermodynamic properties of finite-temperature many-body systems. The results obtained from TPQ calculations converge exponentially in probability toward the exact values as the system size increases \cite{mTPQ, cTPQ, gTPQ}. A key advantage of the TPQ approach over full exact diagonalization (Full-ED) is its reduced memory requirement: instead of storing $O(N)$ eigenvectors for finite-temperature computations, TPQ requires only a single random vector. Since a TPQ state is a pure state in the Hilbert space, nearly all techniques developed for ED—such as symmetry exploitation \cite{Symm} to reduce memory and enhance performance—can be directly applied. Efficient indexing of basis states, for instance via perfect hashing methods \cite{Hash1, Hash2}, remains essential for high performance in such simulations.

    In the original formulation of the TPQ method \cite{cTPQ}, the thermal state is constructed by expanding the matrix exponential via a Taylor series and iteratively applying each term to a random vector until convergence. Although this approach is conceptually simple and easy to implement, it is computationally inefficient and prone to significant numerical errors. In this work, we introduce a practical and high-performance algorithm for generating TPQ states that achieves greater accuracy. The improvement is demonstrated through direct comparisons with results from $\mathcal{H}\Phi$ 3.5.2, confirming both the enhanced precision and efficiency of our method.

    The article is structured as follows. In Section II, we provide a brief overview of the Full-ED and TPQ algorithms, followed by the introduction of our improved method for TPQ state simulation. In Section III, we conduct a comparative analysis of the numerical stability and performance between our proposed algorithm and the original implementation. We conclude with remarks and outlook in Section IV.

\section{Formalism}

\subsection{Full exact diagonalization}

    For a Hamiltonian $\hat H$, the partition function is defined as $Z = \text{tr}[e^{-\beta\hat H}]$, where $\beta = 1/kT$ denotes the inverse temperature. The canonical ensemble average of an observable $\hat O$ is given by $\braket{\hat O}_{\text{ens}} = \frac{1}{Z} \text{tr}[e^{-\beta\hat H} \hat O]$. To evaluate the trace, one may diagonalize the Hamiltonian to obtain its eigenbasis ${\ket{i}}$ and eigenvalues ${E_i}$, allowing the partition function and observable expectation values to be expressed as sums over eigenstates:
    \begin{equation}
        Z = \sum_{i=1}^D e^{-\beta E_i}, \
        \braket{\hat O} = \frac{1}{Z} \sum_{i=1}^D e^{-\beta E_i} \braket{i|\hat O|i},
    \end{equation}
    where $D$ is the dimension of the many-body Hilbert space. This approach constitutes the Full-ED method. However, since $D$ grows exponentially with system size, the $\mathcal{O}(D^2)$ space complexity of Full-ED restricts its practical use to small or toy models.

    A common strategy to mitigate this limitation is to retain only the $M$ lowest-energy eigenstates, which dominate thermodynamic properties at low temperatures. This reduces the space complexity to $\mathcal{O}(MD)$. Nevertheless, the choice of $M$ lacks rigorous convergence guarantees and must be determined empirically. Moreover, even with this truncation, the memory requirement remains substantially higher than that of ground-state-specific methods such as the Lanczos algorithm \cite{Lanczos}.

\subsection{Thermal Quantum Pure states}

    The TPQ states offer an approach to finite-temperature simulations with space complexity comparable to ground-state methods. This is achieved by distributing part of the Hilbert space complexity into the imaginary-time dimension through a stochastic construction. Consider a canonical ensemble at inverse temperature $\beta$ with fixed particle number $N$, and let ${\ket{i}}$ denote the energy eigenstates of the system Hamiltonian $\hat H$. Sampling random complex coefficients ${c_i}$ from the unit hypersphere such that $\sum_i |c_i|^2 = 1$, we define the infinite-temperature canonical TPQ state as $\ket{0} = \sum_i c_i \ket{i}$. The finite-temperature canonical TPQ state is then constructed via imaginary-time evolution:
    \begin{equation}
        \ket{\beta} = e^{-\beta \hat H / 2} \ket{0}.
    \end{equation}

    The expectation value of an observable $\hat O$ is estimated using the expression:
    \begin{equation}
        Z = \overline{\braket{\beta|\beta}}, \
        \braket{\hat O}^{\text{TPQ}} = \frac{1}{Z} \overline{\braket{\beta|\hat O|\beta}},
    \end{equation}
    where $\overline{\cdots}$ indicates the average over random realizations \cite{cTPQ}. It can be shown that for any $\epsilon > 0$, the deviation probability $P\left(\left| \braket{\hat O}^{\text{TPQ}} - \braket{\hat O}_{\text{ens}} \right| > \epsilon\right)$ decays exponentially with increasing system size $N$, where $\braket{\hat O}_{\text{ens}}$ denotes the canonical ensemble average given above.

    To numerically obtain $\ket{\beta}$, the original algorithm expands the matrix exponential as a Taylor series and truncates it at a sufficiently high order $n_0$:
    \begin{equation}
        \ket{\beta} = \sum_{n=0}^\infty \frac{1}{n!} \left(-\frac{\beta \hat H}{2}\right)^n \ket{0}
                    \approx \sum_{n=0}^{n_0} \frac{1}{n!} \left(-\frac{\beta \hat H}{2}\right)^n \ket{0}.
    \end{equation}

\subsection{Improvements on TPQ}

    The original algorithm suffers from several limitations. First, the truncation order $n_0$ is not known a priori and must be determined empirically. Second, even when identified, $n_0$ is typically large due to the slow convergence of the Taylor series expansion. Moreover, numerical overflow may occur due to the rapidly growing terms in the expansion. To address these issues, we adopt the algorithm proposed in \cite{MatExp}, which is suitable for scalable matrix exponential operations in large Hilbert spaces. The matrix exponential is approximated as:
    \begin{equation}
        e^{\bm A} \approx e^{\overline{A}} \left[ T_m\left( \frac{1}{n}(\bm A - \overline{A} \bm I) \right) \right]^n,
        \label{ImagEvo}
    \end{equation}
    where $\bm A = -\frac{\beta}{2} \bm H$, with $\bm H$ being the matrix representation of the Hamiltonian in a chosen basis and $\bm I$ the identity matrix of the same dimension. Here, $\overline{A} = \frac{1}{D} \text{tr}[\bm A]$ is a real scalar that isolates the diagonal contribution to the partition function, thereby enhancing numerical stability. The function $T_m$ denotes the $m$-th order Taylor expansion of the exponential function, and $n$ represents the number of imaginary-time segments. The optimal values of $m$ and $n$ that minimize computational cost while achieving machine precision can be expressed using the 1-norm of matrixes.

    \begin{algorithm}
        \caption{Determination of $(m, n)$} \label{ParamAlgo}
        \KwData{Matrix $\bm A$, $m_\text{max} = 55$, $p_\text{max} = 8$, $\{\theta_i : 1 \le i \le m_\text{max}\}$ as provided in TABLE. 3.1 of \cite{MatExp}}
        \KwResult{Parameters $(m, n)$}

        $N_* \leftarrow 2p_\text{max}(p_\text{max} + 3) \theta_{m_\text{max}}/m_\text{max}$\;
        $N \leftarrow ||\bm A - \overline{A}\bm I||_1$\;
        \eIf{$N < N_*$} {
            $m \leftarrow \text{argmin}_i \; i\lceil N/\theta_i \rceil$\;
            $n \leftarrow \lceil N/\theta_m \rceil$\;
        } {
            Calculate $\{d_p = N \cdot ||\left( \frac{1}{N}(\bm A - \overline{A}\bm I) \right)^p||^{1/p} : 2 \le p \le p_\text{max} + 1\}$\;
            Calculate $\{\alpha_p = \max(d_p, d_{p + 1}) : 2 \le p \le p_\text{max}$\}\;
            Calculate $\{C_i = m\lceil \alpha_p/\theta_i \rceil : 2 \le p \le p_\text{max}, p(p - 1) - 1 \le m \le m_\text{max}\}$\;
            $m \leftarrow \text{argmin}_i \; C_i$\;
            $n \leftarrow \max(\lceil C_m/\theta_m \rceil$, 1)\;
        }
    \end{algorithm}

    In this work, we employ power-based 1-norm estimator \cite{Norm1} rather than the blocked estimator recommended in the original literature, as the latter requires storing multiple eigenstates and BLAS3 matrix-matrix products-a prohibitive requirement for larger systems, whereas the power-based 1-norm estimator only uses BLAS2 matrix-vector product. The matrix polynomial is carefully normalized to avoid overflow(Algorithm \ref{ParamAlgo}). Typically, the matrix-vector product either directly operates on sparse matrix $\bm H$ or is implemented using the on-the-fly trick. However, simply applying these approaches to matrix polynomial $\left( \frac{1}{N}(\bm A - \overline{A}\bm I) \right)^p$ would introduce significant inefficiency. Since modifications to either the large-scale sparse matrix or the on-the-fly Hamiltonian matrix elements have an $O(n^2)$ time complexity, which is generally undesired. Here we incorporate the template expression technique, which lowers the custom matrix-vector product with a standard matrix-vector product and highly optimized BLAS1 operations for arbitrary Hamiltonian without coding effort:
    \begin{equation}
        \frac{1}{N} (\bm A - \overline{A}\bm I)\bm x = -\frac{\beta}{2N}(\bm{Hx}) - \frac{\overline{A}}{N}\bm x.
    \end{equation}
    where $\bm x$ is a given vector. Parameters $(m, n)$ reflects inherent properties of the system's imaginary time evolution of duration $\beta$. As a result, they may be saved and reused for all independent TPQ state simulations, thus avoiding unnecessary computational cost.

    \begin{algorithm}
        \caption{Evaluation of Eq. \ref{ImagEvo}} \label{ImagEvoAlgo}
        \KwData{Matrix $\bm A$, parameters $(m, n)$, arbitrary vector $\bm x$, machine precision $\epsilon$}
        \KwResult{Vector $\bm y = e^{\bm A - \eta \bm I} \bm x$, real scalar $\eta$}

        $i \leftarrow 0$\;
        $\bm y \leftarrow \bm x$\;
        $\eta \leftarrow 0$\;
        \For{$i < n$} {
            $N_\infty \leftarrow ||\bm y||_\infty$\;
            $\bm y \leftarrow \bm y / N_\infty$\;
            $\bm z \leftarrow \bm y$\;
            \For{$j < m$} {
                $\bm z \leftarrow \frac{1}{mj}(\bm A - \overline{A}\bm I)\bm z$\;
                $\bm y \leftarrow \bm y + \bm z$\;
                \If{$N_\infty + ||\bm z||_\infty \le \epsilon ||\bm y||_\infty$} {
                    \textbf{break}\;
                }
                $j \leftarrow j + 1$\;
            }
            $\eta \leftarrow \eta + \ln N_\infty$\;
            $i \leftarrow i + 1$\;
        }
    \end{algorithm}

    Leveraging available symmetries and working within the irreducible Hilbert space is essential for enabling large-scale simulations. Taking the Hubbard model and particle number conservation as example, the Hamiltonian matrix can be diagonalized into a block-diagonal form:
    \begin{equation}
        \bm H = \text{diag}\left( \bm H_0, \bm H_1, \cdots, \bm H_L \right),
    \end{equation}
    where $0, 1, \cdots, L$ are eigenvalues of $\hat N$, $\bm H_l$ is the Hamiltonian matrix in the subspace spaned by states of $l$ particles. One can diagonalize the Hamiltonian in the corresponding subspaces and readily write the imaginary time evolution operator as:
    \begin{equation}
        e^{\bm A} = \text{diag}\left( e^{\bm A_0}, e^{\bm A_1}, \cdots, e^{\bm A_L} \right),
    \end{equation}
    where $\bm A_l = -\frac{\beta}{2}\bm H_l$. The matrix-vector product may be processed in each sectors independently:
    \begin{equation}
        \begin{aligned}
            e^{\bm A}{\bm x} &= \text{diag}\left( e^{\bm A_0}, e^{\bm A_1}, \cdots, e^{\bm A_L} \right){\bm x} \\
                            &= e^{\bm A_0}{\bm x_0} \oplus e^{\bm A_1}{\bm x_1} \oplus \cdots \oplus e^{\bm A_L}{\bm x_L},
        \end{aligned}
    \end{equation}
    
    The expectation value of observable $\hat{O}$ in the grand canonical ensemble is given by:
    \begin{equation}
        \braket{\hat O} = \frac{1}{\Xi} \overline{ \braket{\beta, \mu | \hat{O} | \beta, \mu} },
    \end{equation}
    where $\Xi$ is the grand canonical partition function. The grand canonical TPQ state $\ket{\beta, \mu}$ is related to the canonical TPQ state $\ket{\beta}$ via:
    \begin{equation}
        \ket{\beta, \mu} = e^{\frac{\beta}{2} \mu \hat{N}} \ket{\beta},
    \end{equation}
    with $\mu$ denoting the chemical potential. If the operator $\hat{O}$ commutes with $\hat{N}$, we can further reduce grand canonical expectation values to a weighted sum over symmetry sectors:
    \begin{equation}
        \begin{aligned}
            \braket{\hat O} &= \frac{1}{\Xi} \overline{ \braket{\beta | \hat{O} \, e^{\beta \mu \hat{N}} | \beta} } \\
                            &= \frac{1}{\Xi} \sum_l e^{\beta \mu l} \sum_{\bm{x}_l} \bm{x}_l^\dagger e^{\bm A_l} \bm O e^{\bm A_l} \bm{x}_l,
        \end{aligned}
    \end{equation}
    where $\bm x_l$ is a random initial state in the subspace with fixed particle number $l$. Similarly, the grand canonical partition function can be expressed as:
    \begin{equation}
        \begin{aligned}
            \Xi &= \overline{ \braket{\beta, \mu | \beta, \mu} } \\
                &= \sum_l e^{\beta \mu l} \overline{ \braket{\beta | \beta} }_l,
        \end{aligned}
    \end{equation}
    which requires computing the canonical partition function $Z_l = \overline{ \braket{\beta | \beta} }_l = \sum_{\bm{x}_l} |e^{\bm A_l} \bm{x}_l|^2$. Note that partition function scales exponentially with system size and inverse temperature $\beta$, but acts as a normalization factor of TPQ states and does not affect the essential physics. Therefore, in practice, we accumulate the logarithmic partition function for each imaginary time step separately, rather than explicitly forming $e^{\bm A_l} \bm x$. This approach is critical for reaching arbitrary low temperatures while keeping enough effective digits(Algorithm \ref{ImagEvoAlgo}). We have implemented above algorithms in our open-source C++ template library, \textit{Physica} (Appendix \ref{HighLevelDesign}). To keep the article tight, we have provided guidance on reproducing our results in Appendix \ref{Guidance}.

\section{Results and discussions}

    We use the Hubbard model \cite{Hubbard} as our test case, which is widely regarded as the minimal two-dimensional (2D) model for cuprate superconductors. The Hamiltonian, including nearest-neighbor hopping, is given by:
    \begin{equation}
        \hat H = -\sum_{ij, \sigma} t_{ij} \left( \hat c_{i\sigma}^\dagger \hat c_{j\sigma} + \text{h.c.} \right)
               + U \sum_i \hat n_{i \uparrow} \hat n_{i \downarrow} - \mu \sum_i \left( \hat n_{i \uparrow} + \hat n_{i \downarrow} \right)
    \end{equation}
    where $\hat c_{i\sigma}^\dagger$, $\hat c_{i\sigma}$, and $\hat n_{i\sigma}$ denote the creation, annihilation, and particle number operators, respectively, for an electron at site $i$ with spin $\sigma$ ($\uparrow$ or $\downarrow$). The hopping parameter $t_{ij}$ is equal to $t$ for nearest-neighbor pairs and zero otherwise. We consider a strong interaction strength of $U/t = 8$ and simulate multiple systems under periodic boundary conditions (PBC).

    \subsection{Numerical stability}

    \begin{figure}
        \refstepcounter{figure} \label{Stability}
        \begin{center} 
            \def\svgwidth{0.4\textwidth}
            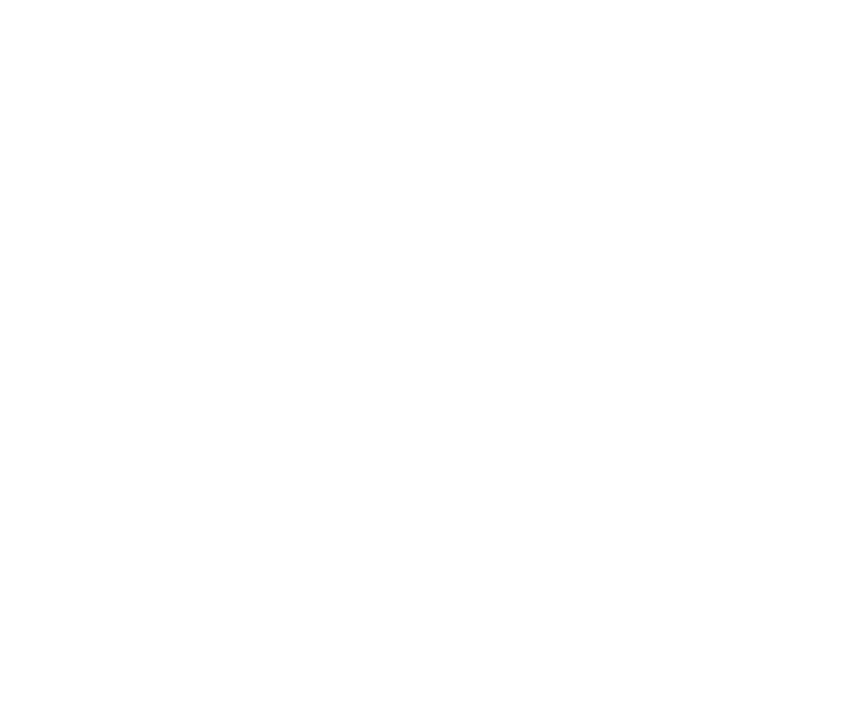
            \parbox[c]{15.0cm}{\footnotesize{\bf Fig.~1.}
            The relative error $R$ in the electron density for the one-dimensional 4-site Hubbard model at $\mu = 0$, showing its dependence on inverse temperature $\beta$ for our improved algorithm(red solid line) and the original algorithm implemented in $\mathcal{H}\Phi$(blue solid line). The statistical uncertainty($2\sigma_R$) for each is indicated by the lighter shaded regions. Results are benchmarked against the Full-ED results from $\mathcal{H}\Phi$(black circles). TPQ expectation values are computed using a combination of $8$ independent runs and $2^{13} = 8192$ random initial states per run. The imaginary time step size is taken to be $\Delta\tau = 0.1$. }
        \end{center}
    \end{figure}

    To validate the correctness of our implementation, we begin by studying the one-dimensional Hubbard model with 4 sites at $\mu = 0$—a system size amenable to Full-ED. The electron density $\rho$ as a function of inverse temperature $\beta$ is computed using TPQ methods implemented independently in our framework and in $\mathcal{H}\Phi$. Using the Full-ED results obtained from $\mathcal{H}\Phi$ as the reference, the relative error $R$ and its corresponding standard deviation $\sigma_R$ for the electron density is defined as:
    \begin{equation}
        R = \frac{\rho - \rho_{\text{Full-ED}}}{\rho_{\text{Full-ED}}}, \
        \sigma_R = \frac{\sigma_\rho}{\rho_{\text{Full-ED}}},
    \end{equation}
    where $\sigma_\rho$ is standard deviation of electron density.

    As shown in \cref{Stability}, our TPQ results demonstrate excellent agreement with Full-ED across the temperature range. We further note that the initial states are not necessarily generated in complex number space, but satisfying the uncorrelated phase condition $\braket{c_i c_j} = \delta_{ij}/D$ is sufficient to allow destructive interference between different eigenstates. Thus, the statistical uncertainty is much smaller due to the selection of real number initial states. In contrast, the original TPQ algorithm implemented in $\mathcal{H}\Phi$ exhibits rapidly growing relative error, reaching approximately $3\%$ as temperature decreases, highlighting its numerical instability at low temperatures. While increasing the expansion order in $\mathcal{H}\Phi$ can improve accuracy, the associated computational cost becomes prohibitively high, especially for larger systems. As a result, we attribute this error accumulation to the original algorithms' use of a non-adaptive truncation of the expansion order. The eventual decrease in relative error at very low temperatures occurs as the system approaches its ground state, where contributions from high energy are less significant.

    \subsection{Performance}
    
    We simulate the Hubbard model on a $4 \times 4$ square lattice, which, to the best of our knowledge, still lacks exact numerical results across arbitrary temperatures and doping levels. The simulations are carried out at multiple temperatures, extending down to $\beta t = 32$. The system is hole-doped, with the chemical potential $\mu/t$ varied over the interval $[-8, 0]$, a range sufficiently broad to encompass regimes from Fermi liquids to potential superconducting phases (\cref{4x4}). Taking the $\beta t = 16$ curve as an example, we compute full profiles of electron density and double occupancy as functions of chemical potential in approximately $5 \times 10^3$ core-hours on our test platform (Intel$\circledR$ Xeon$\circledR$ Platinum 8358, 256 GB RAM). In contrast, under equivalent computational conditions, $\mathcal{H}\Phi$ produces only a limited number of data points; we estimate that generating the complete curve would require approximately $10^6$ core-hours.

    \begin{center} 
        \refstepcounter{figure} \label{4x4}
        \def\svgwidth{0.4\textwidth}
\begingroup%
  \makeatletter%
  \providecommand\color[2][]{%
    \errmessage{(Inkscape) Color is used for the text in Inkscape, but the package 'color.sty' is not loaded}%
    \renewcommand\color[2][]{}%
  }%
  \providecommand\transparent[1]{%
    \errmessage{(Inkscape) Transparency is used (non-zero) for the text in Inkscape, but the package 'transparent.sty' is not loaded}%
    \renewcommand\transparent[1]{}%
  }%
  \providecommand\rotatebox[2]{#2}%
  \newcommand*\fsize{\dimexpr\f@size pt\relax}%
  \newcommand*\lineheight[1]{\fontsize{\fsize}{#1\fsize}\selectfont}%
  \ifx\svgwidth\undefined%
    \setlength{\unitlength}{393.22499084bp}%
    \ifx\svgscale\undefined%
      \relax%
    \else%
      \setlength{\unitlength}{\unitlength * \real{\svgscale}}%
    \fi%
  \else%
    \setlength{\unitlength}{\svgwidth}%
  \fi%
  \global\let\svgwidth\undefined%
  \global\let\svgscale\undefined%
  \makeatother%
  \begin{picture}(1,0.86076673)%
    \lineheight{1}%
    \setlength\tabcolsep{0pt}%
    \put(0,0){\includegraphics[width=\unitlength,page=1]{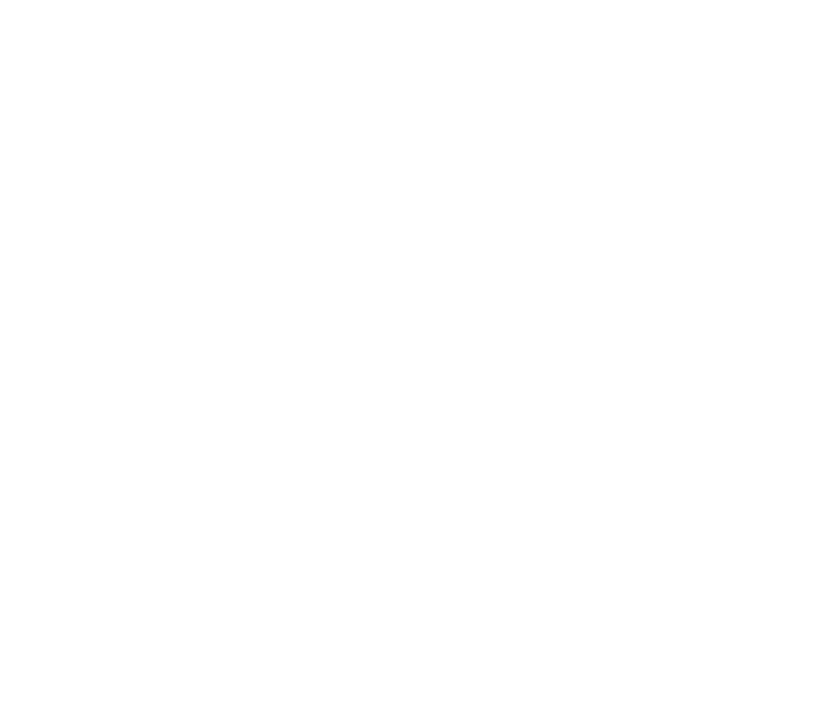}}%
    \put(0.52776751,0.01505701){\color[rgb]{0.25098039,0.25098039,0.26666667}\makebox(0,0)[lt]{\lineheight{1.25}\smash{\begin{tabular}[t]{l}$\mu$\end{tabular}}}}%
    \put(0.52776751,0.88505701){\color[rgb]{0.25098039,0.25098039,0.26666667}\makebox(0,0)[lt]{\lineheight{1.25}\smash{\begin{tabular}[t]{l}(a)\end{tabular}}}}%
    \put(0.02333092,0.46974078){\color[rgb]{0.25098039,0.25098039,0.26666667}\rotatebox{90}{\makebox(0,0)[lt]{\lineheight{1.25}\smash{\begin{tabular}[t]{l}$\rho$\end{tabular}}}}}%
    \put(0,0){\includegraphics[width=\unitlength,page=2]{BetaRho.pdf}}%
    \put(0.09047397,0.07036885){\color[rgb]{0.25098039,0.25098039,0.26666667}\makebox(0,0)[lt]{\lineheight{1.25}\smash{\begin{tabular}[t]{l}-8\end{tabular}}}}%
    \put(0.30599943,0.07036885){\color[rgb]{0.25098039,0.25098039,0.26666667}\makebox(0,0)[lt]{\lineheight{1.25}\smash{\begin{tabular}[t]{l}-6\end{tabular}}}}%
    \put(0.52343221,0.07036885){\color[rgb]{0.25098039,0.25098039,0.26666667}\makebox(0,0)[lt]{\lineheight{1.25}\smash{\begin{tabular}[t]{l}-4\end{tabular}}}}%
    \put(0.73895768,0.07036885){\color[rgb]{0.25098039,0.25098039,0.26666667}\makebox(0,0)[lt]{\lineheight{1.25}\smash{\begin{tabular}[t]{l}-2\end{tabular}}}}%
    \put(0.96020506,0.07036885){\color[rgb]{0.25098039,0.25098039,0.26666667}\makebox(0,0)[lt]{\lineheight{1.25}\smash{\begin{tabular}[t]{l}0\end{tabular}}}}%
    \put(0,0){\includegraphics[width=\unitlength,page=3]{BetaRho.pdf}}%
    \put(0.03469865,0.10279304){\color[rgb]{0.25098039,0.25098039,0.26666667}\makebox(0,0)[lt]{\lineheight{1.25}\smash{\begin{tabular}[t]{l}0.0\end{tabular}}}}%
    \put(0.03469865,0.2420263){\color[rgb]{0.25098039,0.25098039,0.26666667}\makebox(0,0)[lt]{\lineheight{1.25}\smash{\begin{tabular}[t]{l}0.2\end{tabular}}}}%
    \put(0.03469865,0.38125957){\color[rgb]{0.25098039,0.25098039,0.26666667}\makebox(0,0)[lt]{\lineheight{1.25}\smash{\begin{tabular}[t]{l}0.4\end{tabular}}}}%
    \put(0.03469865,0.52049284){\color[rgb]{0.25098039,0.25098039,0.26666667}\makebox(0,0)[lt]{\lineheight{1.25}\smash{\begin{tabular}[t]{l}0.6\end{tabular}}}}%
    \put(0.03469865,0.6597261){\color[rgb]{0.25098039,0.25098039,0.26666667}\makebox(0,0)[lt]{\lineheight{1.25}\smash{\begin{tabular}[t]{l}0.8\end{tabular}}}}%
    \put(0.03469865,0.79895937){\color[rgb]{0.25098039,0.25098039,0.26666667}\makebox(0,0)[lt]{\lineheight{1.25}\smash{\begin{tabular}[t]{l}1.0\end{tabular}}}}%
    \put(0,0){\includegraphics[width=\unitlength,page=4]{BetaRho.pdf}}%
    \put(0.3561318,0.76939614){\color[rgb]{0.25098039,0.25098039,0.26666667}\makebox(0,0)[lt]{\lineheight{1.25}\smash{\begin{tabular}[t]{l}$\beta=2$\end{tabular}}}}%
    \put(0,0){\includegraphics[width=\unitlength,page=5]{BetaRho.pdf}}%
    \put(0.3561318,0.70836238){\color[rgb]{0.25098039,0.25098039,0.26666667}\makebox(0,0)[lt]{\lineheight{1.25}\smash{\begin{tabular}[t]{l}$\beta=4$\end{tabular}}}}%
    \put(0,0){\includegraphics[width=\unitlength,page=6]{BetaRho.pdf}}%
    \put(0.3561318,0.64732862){\color[rgb]{0.25098039,0.25098039,0.26666667}\makebox(0,0)[lt]{\lineheight{1.25}\smash{\begin{tabular}[t]{l}$\beta=8$\end{tabular}}}}%
    \put(0,0){\includegraphics[width=\unitlength,page=7]{BetaRho.pdf}}%
    \put(0.3561318,0.58629486){\color[rgb]{0.25098039,0.25098039,0.26666667}\makebox(0,0)[lt]{\lineheight{1.25}\smash{\begin{tabular}[t]{l}$\beta=16$\end{tabular}}}}%
    \put(0,0){\includegraphics[width=\unitlength,page=8]{BetaRho.pdf}}%
    \put(0.3561318,0.5252611){\color[rgb]{0.25098039,0.25098039,0.26666667}\makebox(0,0)[lt]{\lineheight{1.25}\smash{\begin{tabular}[t]{l}$\beta=32$\end{tabular}}}}%
    \put(0,0){\includegraphics[width=\unitlength,page=9]{BetaRho.pdf}}%
  \end{picture}%
\endgroup%

        \def\svgwidth{0.4\textwidth}
\begingroup%
  \makeatletter%
  \providecommand\color[2][]{%
    \errmessage{(Inkscape) Color is used for the text in Inkscape, but the package 'color.sty' is not loaded}%
    \renewcommand\color[2][]{}%
  }%
  \providecommand\transparent[1]{%
    \errmessage{(Inkscape) Transparency is used (non-zero) for the text in Inkscape, but the package 'transparent.sty' is not loaded}%
    \renewcommand\transparent[1]{}%
  }%
  \providecommand\rotatebox[2]{#2}%
  \newcommand*\fsize{\dimexpr\f@size pt\relax}%
  \newcommand*\lineheight[1]{\fontsize{\fsize}{#1\fsize}\selectfont}%
  \ifx\svgwidth\undefined%
    \setlength{\unitlength}{390.75bp}%
    \ifx\svgscale\undefined%
      \relax%
    \else%
      \setlength{\unitlength}{\unitlength * \real{\svgscale}}%
    \fi%
  \else%
    \setlength{\unitlength}{\svgwidth}%
  \fi%
  \global\let\svgwidth\undefined%
  \global\let\svgscale\undefined%
  \makeatother%
  \begin{picture}(1,0.86948177)%
    \lineheight{1}%
    \setlength\tabcolsep{0pt}%
    \put(0.54588618,0.01382553){\color[rgb]{0.25098039,0.25098039,0.26666667}\makebox(0,0)[lt]{\lineheight{1.25}\smash{\begin{tabular}[t]{l}$\mu$\end{tabular}}}}%
    \put(0.54588618,0.88382553){\color[rgb]{0.25098039,0.25098039,0.26666667}\makebox(0,0)[lt]{\lineheight{1.25}\smash{\begin{tabular}[t]{l}(b)\end{tabular}}}}%
    \put(0.02648292,0.42835298){\color[rgb]{0.25098039,0.25098039,0.26666667}\rotatebox{90}{\makebox(0,0)[lt]{\lineheight{1.25}\smash{\begin{tabular}[t]{l}$\braket{\hat n_\uparrow \hat n_\downarrow}$\end{tabular}}}}}%
    \put(0,0){\includegraphics[width=\unitlength,page=1]{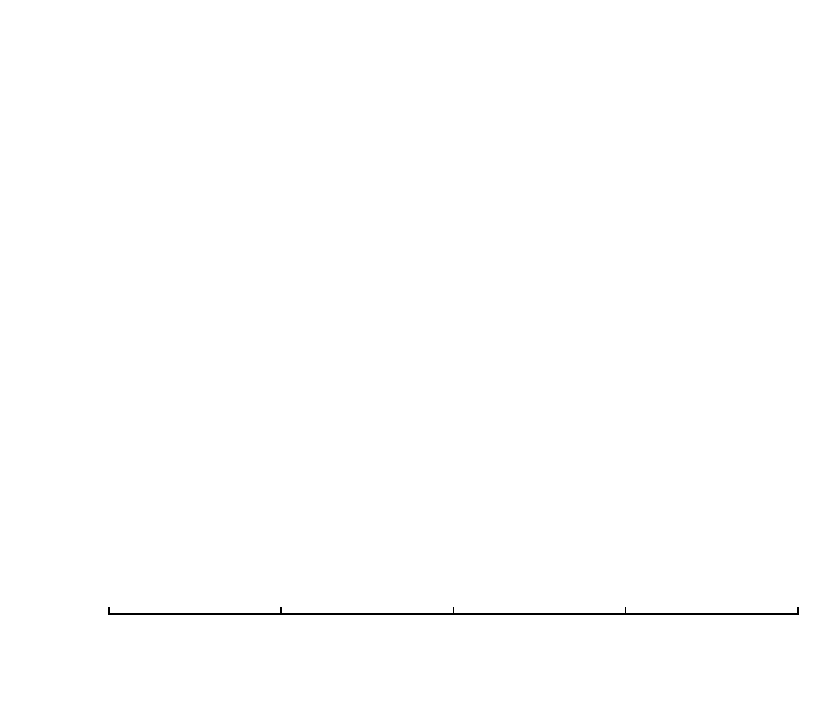}}%
    \put(0.11708388,0.06948772){\color[rgb]{0.25098039,0.25098039,0.26666667}\makebox(0,0)[lt]{\lineheight{1.25}\smash{\begin{tabular}[t]{l}-8\end{tabular}}}}%
    \put(0.32821631,0.06948772){\color[rgb]{0.25098039,0.25098039,0.26666667}\makebox(0,0)[lt]{\lineheight{1.25}\smash{\begin{tabular}[t]{l}-6\end{tabular}}}}%
    \put(0.54126814,0.06948772){\color[rgb]{0.25098039,0.25098039,0.26666667}\makebox(0,0)[lt]{\lineheight{1.25}\smash{\begin{tabular}[t]{l}-4\end{tabular}}}}%
    \put(0.75240058,0.06948772){\color[rgb]{0.25098039,0.25098039,0.26666667}\makebox(0,0)[lt]{\lineheight{1.25}\smash{\begin{tabular}[t]{l}-2\end{tabular}}}}%
    \put(0.96929117,0.06948772){\color[rgb]{0.25098039,0.25098039,0.26666667}\makebox(0,0)[lt]{\lineheight{1.25}\smash{\begin{tabular}[t]{l}0\end{tabular}}}}%
    \put(0,0){\includegraphics[width=\unitlength,page=2]{BetaD.pdf}}%
    \put(0.03498599,0.10211727){\color[rgb]{0.25098039,0.25098039,0.26666667}\makebox(0,0)[lt]{\lineheight{1.25}\smash{\begin{tabular}[t]{l}0.00\end{tabular}}}}%
    \put(0.03498599,0.34779866){\color[rgb]{0.25098039,0.25098039,0.26666667}\makebox(0,0)[lt]{\lineheight{1.25}\smash{\begin{tabular}[t]{l}0.02\end{tabular}}}}%
    \put(0.03498599,0.59156065){\color[rgb]{0.25098039,0.25098039,0.26666667}\makebox(0,0)[lt]{\lineheight{1.25}\smash{\begin{tabular}[t]{l}0.04\end{tabular}}}}%
    \put(0.03498599,0.83724203){\color[rgb]{0.25098039,0.25098039,0.26666667}\makebox(0,0)[lt]{\lineheight{1.25}\smash{\begin{tabular}[t]{l}0.06\end{tabular}}}}%
    \put(0,0){\includegraphics[width=\unitlength,page=3]{BetaD.pdf}}%
    \put(0.41299405,0.77292572){\color[rgb]{0.25098039,0.25098039,0.26666667}\makebox(0,0)[lt]{\lineheight{1.25}\smash{\begin{tabular}[t]{l}$\beta=2$\end{tabular}}}}%
    \put(0,0){\includegraphics[width=\unitlength,page=4]{BetaD.pdf}}%
    \put(0.41299405,0.71150557){\color[rgb]{0.25098039,0.25098039,0.26666667}\makebox(0,0)[lt]{\lineheight{1.25}\smash{\begin{tabular}[t]{l}$\beta=4$\end{tabular}}}}%
    \put(0,0){\includegraphics[width=\unitlength,page=5]{BetaD.pdf}}%
    \put(0.41299405,0.65008522){\color[rgb]{0.25098039,0.25098039,0.26666667}\makebox(0,0)[lt]{\lineheight{1.25}\smash{\begin{tabular}[t]{l}$\beta=8$\end{tabular}}}}%
    \put(0,0){\includegraphics[width=\unitlength,page=6]{BetaD.pdf}}%
    \put(0.41299405,0.58866488){\color[rgb]{0.25098039,0.25098039,0.26666667}\makebox(0,0)[lt]{\lineheight{1.25}\smash{\begin{tabular}[t]{l}$\beta=16$\end{tabular}}}}%
    \put(0,0){\includegraphics[width=\unitlength,page=7]{BetaD.pdf}}%
    \put(0.41299405,0.52724453){\color[rgb]{0.25098039,0.25098039,0.26666667}\makebox(0,0)[lt]{\lineheight{1.25}\smash{\begin{tabular}[t]{l}$\beta=32$\end{tabular}}}}%
    \put(0,0){\includegraphics[width=\unitlength,page=8]{BetaD.pdf}}%
  \end{picture}%
\endgroup%

        \parbox[c]{15.0cm}{\footnotesize{\bf Fig.~2.}
        (a) Electron density ($\rho$) and (b) double occupation ($\braket{\hat n_\uparrow \hat n_\downarrow}$) as a function of chemical potential ($\mu$) for a $4 \times 4$ Hubbard model with PBC. Results are plotted for several inverse temperatures: $\beta = 2$(blue), $4$(orange), $8$(gold), $16$(purple) and $32$(green).}
    \end{center}

\section{Conclusion}

    We present a high-performance and numerically stable algorithm for simulating TPQ states, rendering the TPQ approach highly practical for solving quantum many-body problems. This algorithm is implemented within an open-source C++ template library, combining computational efficiency with robust numerical reliability. Our improved method demonstrates significantly better agreement with exact numerical benchmarks compared to conventional approaches. Moreover, the implementation achieves a speedup of three orders of magnitude over existing methods in $\mathcal{H}\Phi$. By extending the limits of scalable and accurate benchmarking in quantum many-body systems, this work provides a solid and efficient foundation for validating emerging computational algorithms in the field.

\addcontentsline{toc}{chapter}{Acknowledgment}
\section*{Acknowledgment}
    The authors acknowledge Fu-Zhou Chen for helpful discussions. The work is partly supported by the National Key Research and Development Program of China (Grant No. 2022YFA1402704) and the programs for NSFC of China (Grant No. 12247101).

\begin{appendices}
    \section{High level design of \textit{Physica} \label{HighLevelDesign}}
        \subsection*{1. Project structure}

        The project structure is often the first aspect users encounter when engaging with open-source software. A clean and goal-oriented project organization is as crucial as well-designed and well-documented code. The top-level directory structure of \textit{Physica} follows conventions commonly adopted by open-source C++ projects, consisting of the following directories: \verb|3rdparty|, \verb|benchmark|, \verb|doc|, \verb|examples|, \verb|include|, \verb|src|, and \verb|test|. Within \textit{Physica}, the secondary structure is organized modularly. The modules included in \textit{Physica} are listed in \cref{Modules}.

        \begin{center}
            \refstepcounter{table} \label{Modules}
            \footnotesize{\bf Table 1.} Modules and description.\\
            \vspace{2mm}
            \begin{tabular}{ccc}
                \hline
                Module & Description \\
                \hline
                Core & Implementation of Physica's core functionality \\
                Gui & Includes 2D and 3D drawing support, using Qt as the drawing backend \\
                Logger & A high-performance logging library based on NanoLog\cite{NanoLog} \\
                Python & Backend of Physica python interface(WIP) \\
                phypy & Physica python interface(WIP) \\
                \hline
            \end{tabular}
        \end{center}

        The secondary project structure extends to each directory within the top-level layout. Tertiary and finer-grained structures encompass APIs and implementation details. The organization follows scientific—rather than purely engineering—logic, allowing domain experts to more readily adopt \textit{Physica}. Engineering complexities are encapsulated within directories suffixed with ``Impl'', with lower-level logic nested deeper in the directory hierarchy.

        Overall, \textit{Physica} employs a goal-oriented, layered structure that progressively exposes complexity. Users not concerned with implementation details will generally encounter fewer such details, as most common use cases are addressed at shallower directory levels. Since scientific workflows are diverse and often require flexibility, we ensure users retain the ability to access and modify underlying implementation details. The pervasive use of templates further facilitates non-intrusive customization and extension.

        \subsection*{2. The rational of templates}

        The development of scalable programs presents significant challenges, as each problem exhibits both universal and particular aspects. A central difficulty lies in balancing these dimensions: generic implementations may require adaptation to improve performance or insight in specific contexts, while optimizations tailored to one system may not transfer effectively to others. In essence, each case demands its own optimally suited implementation. Manually developing specialized solutions for every scenario, however, is often impractical and inefficient.

        To address this, we employ template metaprogramming techniques, which allow code generation rules to instruct the compiler to automatically produce optimized implementations for each use case. Conventional scientific computing programs face a fundamental trade-off: they often struggle to incorporate problem-specific optimizations without sacrificing generality. Efforts to introduce specificity frequently lead to uncontrolled growth in input and output configurations, resulting in substantial maintenance and efficiency costs.

        Consider, for example, floating-point numeric types: common options include \verb|bfloat16|, \verb|float16|, \verb|float32|, \verb|float64|, and \verb|float128|. When extended to complex numbers, the number of available types doubles. Incorporating automatic differentiation—as often required in deep learning—further doubles this number, leading to at least 20 possible floating-point type combinations, even before considering future extensions. Without templates, each function would require over 20 separate implementations, resulting in repetitive and hard-to-maintain code. By contrast, with C++ template metaprogramming, all valid type combinations are resolved automatically at compile time. Since template parameters are evaluated during compilation, this approach ensures both performance and extensibility, readily accommodating new numerical types without code modification.

        However, excessive reliance on templates can lead to long compilation times, large binary sizes, and increased complexity. To mitigate these issues, we adopt the following strategies:

        1. Although \textit{Physica} is primarily header-based—a natural consequence of heavy template use that also simplifies usage and modification—we compile sufficiently general and performance-insensitive modules (such as exceptions, I/O, and common utilities) into dynamic libraries. This reduces executable size and compilation time while preserving the benefits of templates where they matter most.

        2. To lower the barrier to entry, we encourage users to consult the example cases provided in the \verb|examples| directory. These serve as practical prototypes that can be adapted as needed and will be continuously updated based on community feedback.

        3. The expressive power of templates enables the formal composition of existing features and straightforward extension to new functionality, supporting both generality and specificity without invasive code changes.

        \subsection*{3. Input and Output}

        \textit{Physica} intentionally does not provide traditional input or output files—a deliberate design choice that reflects its modern approach to scientific computing. As software functionalities expand, the rigid input-compute-output model of the past has become increasingly inadequate for contemporary research needs. Many large-scale scientific computing packages rely on cumbersome input files with hundreds of keywords \cite{LAMMPS, CP2K}, where each new feature introduces additional tags that complicate usage and obscure intent. Like comments and documentation, input files are inherently decoupled from the code, creating a risk of silent discrepancies between user intent and actual computation. Moreover, the limited expressiveness of input files restricts users' ability to finely customize functionality.

        Output files in traditional frameworks also present challenges: it is common for large-scale programs to generate numerous lengthy files, forcing users to navigate extensive irrelevant data to locate meaningful results. This not only impedes productivity but also consumes substantial computational resources for superfluous outputs. Simply adding more tags to control output would lead to a “tag explosion,” further complicating the input specification without solving the underlying inflexibility.

        To address these issues, \textit{Physica} adopts an object-oriented design in which each computational problem is represented as an object. Input parameters are hardcoded into the program to form computational objects, enabling compile-time optimization and ensuring that only relevant objects are constructed—effectively avoiding tag proliferation. This embedded input approach guarantees “what you see is what you compute,” eliminating any disconnect between specification and execution. Furthermore, using a general-purpose programming language offers significantly greater expressivity than domain-specific languages (DSLs), allowing more nuanced and precise descriptions of physical processes.

        For data I/O, \textit{Physica} uses HDF5 as its standard format. Inspired by Unix philosophy, the API treats every computational object as readable and writable. Through object-oriented composition, simple computational objects can be encapsulated into more complex structures. Users can selectively compute and output only the objects—or sub-objects—they need, avoiding unnecessary data generation and streamlining result analysis.

    \section{Usage guide \label{Guidance}}

        \textit{Physica} is an open-source C++ template library distributed under the GNU General Public License version 3. The source code is publicly accessible on Gitee \cite{Physica}. For a detailed installation guide, users may refer to the official documentation \cite{PhysicaDoc}. In the following, we introduce several core concepts of \textit{Physica} before outlining the steps required to reproduce the results presented in \cref{Stability}. The complete source code, along with additional usage examples, is provided in the accompanying examples folder.

        \subsection*{1. Scalar and linear algebra}
        
        \textit{Physica} provides a comprehensive implementation of scalar algebra, and further extends this foundation with robust support for differentiable linear algebra \cite{MatrixCompute}, making it suitable for general-purpose scientific computing. Both scalar and linear algebra components are systematically unified using C++20 concepts. Drawing inspiration from established C++ linear algebra libraries such as Eigen \cite{Eigen} and Armadillo \cite{Armadillo}, the linear algebra module makes extensive use of template expression techniques to eliminate unnecessary temporary objects and enable compile-time expression optimization. SIMD (Single Instruction Multiple Data) intrinsics \cite{VectorClass} are also heavily utilized to improve instruction-level parallelism. Additionally, users can interface with high-performance vendor libraries such as OneMKL \cite{MKL} and CUDA \cite{CUDA} to further accelerate linear algebra operations and enable GPU offloading.

        To begin a simulation, the first step is to select an appropriate scalar type. The real number module, accessible through the corresponding header, provides three floating-point types: \verb|float16|, \verb|float32|, and \verb|float64|. For reasons of numerical stability, we use \verb|float64| throughout this work. A \verb|using namespace| declaration can be employed to conveniently expose these types in the current scope.

        \begin{lstlisting}
#include "Physica/Core/Scalar/Real.h"

using namespace Physica;

float64 HoppingT = 1;
float64 RepelU = 8;
        \end{lstlisting}

        We are interested in studying the electron density $\rho$ as a function of inverse temperature $\beta$. From a numerical perspective, we discretize the imaginary-time axis into multiple intervals and store the corresponding values in an $N$-dimensional vector. The application programming interface (API) is designed to be consistent with conventions in both NumPy \cite{numpy} and MATLAB \cite{MATLAB}. Starting from an infinite-temperature random quantum state, we gradually “cool” the system to a sufficiently low temperature. In this example, the stopping condition is set to $\beta t = 4$, and the imaginary-time domain is discretized into \verb|40| slices.

        \begin{lstlisting}
#include "Physica/Core/Math/Algebra/LinearAlgebra/Vector/DenseVector.h"

int NumBeta = 40 + 1; // Both 0 and 4 are included
auto betas = VectorND<float64>::linspace(0, 4, NumBeta);
        \end{lstlisting}

        \subsection*{2. Modeling of many-body problem}

        In \textit{Physica}, any many-body problem is modeled through three core concepts: \verb|Representation|, \verb|State|, and \verb|Hamiltonian|. The \verb|Representation| class defines a mapping between elements of the Hilbert space and numerical indices. Currently, the library provides two primary representations: \verb|FermiRepr| for fermionic systems and \verb|SpinRepr| for spin systems. Elements of the Hilbert space are represented as \verb|State| objects, with \verb|FermiState| and \verb|SpinState| corresponding to their respective representations. Efficient indexing of basis configurations is essential for performance in ED type algorithms. To this end, we employ a hash table-based indexing mechanism that ensures $O(1)$ time complexity for state lookups. A Hilbert space can be instantiated by constructing a representation object:

        \begin{lstlisting}
#include "Physica/Core/Physics/ManyBody/ReprSpace/FermiRepr.h"

using ReprType = FermiRepr<Dim, NumSite, UseInversionSymm>;
ReprType repr(numSpinUp, numSpinDown);
        \end{lstlisting}

        The \verb|FermiRepr| representation is implemented as a template class with three template parameters: the system's dimension, the total number of sites, and a Boolean flag indicating whether to employ inversion symmetry. When \verb|UseInversionSymm| is set to \verb|true| and the system satisfies \verb|numSpinUp == numSpinDown|, memory usage can be reduced by approximately half. States belonging to the representation are automatically generated during object construction.

        Once the representation object is constructed, the next step is to define the parameters of the Hamiltonian, including the lattice geometry, boundary conditions, interaction strengths, and other relevant terms. The template class \verb|SquareLattice| accepts one template parameter specifying the spatial dimension of the system. In this example, we consider a one-dimensional Hubbard model with \verb|NumSite| sites and one site per unit cell. We can construct the corresponding Hamiltonian in the given representation:

        \begin{lstlisting}
#include "Physica/Core/Physics/ManyBody/Hamilton/HubbardMatrix.h"

using Hamiltonian = HubbardMatrix<float64, ReprType>;
SquareLattice<Dim> lattice({NumSite}, 1);
Hamiltonian H(HoppingT, RepelU, lattice, repr);
        \end{lstlisting}
        where we define the precision of the Hamiltonian matrix as \verb|float64|. The resulting Hamiltonian matrix may be calculated on-the-fly or stored in a sparse matrix, which is provided by our linear algebra submodule.

        \subsection*{3. Simulation of TPQ state}

        Since states under a given representation can be naturally modeled as continuous vectors, the \verb|TPQ| class is inherited from the \textit{n}-dimensional vector and can be accessed like normal vectors. We construct an infinity temperature TPQ state with the size of the Hilbert space and initialize it with gaussian random numbers:
        \begin{lstlisting}
#include "Physica/Core/Physics/ManyBody/TPQ.h"

TPQ<float64> psi(H.getNumState());
psi.random_normal<Random<>>();
        \end{lstlisting}
        where \verb|Random<>| is the Mersenne Twister pseudo-random generator. The imaginary time evolution can be carried out as easy as one line of code:
        \begin{lstlisting}
psi.nvt_step<Hamiltonian>(H, deltaT);
        \end{lstlisting}
        where we evolve the TPQ state by an imaginary time of \verb|deltaT|.

        \subsection*{4. Writing results to HDF5}

        We adopt HDF5 (Hierarchical Data Format version 5) \cite{HDF5} as the standard output format for numerical simulation results. Every C++ class in \textit{Physica} provides two straightforward member functions, \verb|read| and \verb|write|, which facilitate loading data from and saving data to HDF5 files, respectively.

        \begin{lstlisting}
T data{};
auto h5f = H5File::open("data.h5");
data.read(h5f, "x");
// Operations on data...
data.write(h5f, "x");
        \end{lstlisting}
        
        For any data of type \verb|T|, an HDF5 file can be created or opened by calling the static member function \verb|open| from the class \verb|H5File|. Here, the HDF5 file is named `data.h5'. We then read the dataset labeled `x' into memory, perform the necessary processing, and ultimately write the updated data back to the file.

        \subsection*{5. Data Visualization}

        We provide native data visualization capabilities using Qt \cite{Qt} as the backend, supporting a wide range of 2D plots as well as basic 3D visualizations.

        \begin{lstlisting}
#include "Physica/Gui/Plot/Plot.h"

QApplication app(argc, argv);
Plot* plot = new Plot(-5, 5, -1.1, 1.1, 2, 0.5);
auto x = VectorND<float64>::linspace(-5, 5, 100);
auto y = tanh(x);
plot->spline(x, y);
plot->show();
QApplication::exec();
        \end{lstlisting}

        It is essential to initialize Qt by constructing a \verb|QApplication| object before performing any plotting operations. The constructor of \verb|QApplication| accepts the command-line arguments from the \verb|main| function of a standard C++ program. Once Qt is initialized, a \verb|Plot| object can be created to handle 2D plotting tasks. Its constructor accepts six parameters: the minimum and maximum values of the x-axis, the minimum and maximum values of the y-axis, and the step sizes for the x-axis and y-axis. In the example above, we plot the \verb|tanh| function from $-5$ to $5$. The x-axis range is discretized into 100 equal intervals, and the curve between data points is interpolated using a B-spline algorithm. The plotting process is finalized by invoking the \verb|show()| member function of the \verb|Plot| class, and the Qt backend is instructed to render the plot by calling \verb|QApplication::exec()|.

        \subsection*{6. Debugging}
        
        We provide a comprehensive assertion mechanism to assist with error checking, comprising both static and dynamic assertions:

        \textbullet \; Static assertions are used to detect errors in template parameter usage. Evaluated at compile time, they allow users to identify issues early and impose no runtime performance overhead. Static assertions are always enabled.

        \textbullet \; Dynamic assertions are employed where static checks are insufficient—only during runtime. These assertions incur a runtime performance cost and are enabled only in Debug mode to minimize overhead in production builds. Users can compile their program in debug mode or define the macro \verb|NDEBUG| to control the enabling of dynamic assertions.

        This two-tiered approach ensures robust error detection while maintaining high runtime efficiency in release builds.
\end{appendices}

\end{document}